\documentclass[11pt]{article}
\usepackage{latexsym}
\usepackage{epsfig}

\begin{document}

\title{Network of Econophysicists: a weighted network to investigate the development of Econophysics}

\author{Ying Fan$^1$, Menghui Li$^1$, Jiawei Chen$^1$, Liang Gao$^1$, Zengru Di$^1$, Jinshan Wu$^{1,2}${\footnote{Author for correspondence: jinshanw@sfu.edu}}
\\1. Department of Systems Science, School of Management, Beijing Normal University, \\
Beijing, 100875, P.R. of China
\\2. Department of Physics, Simon Fraser University, Burnaby, B.C.
Canada, V5A 1S6}

\maketitle

\begin{abstract}
The development of Econophysics is studied from the perspective of
scientific communication networks. Papers in Econophysics
published from 1992 to 2003 are collected. Then a weighted and
directed network of scientific communication, including
collaboration, citation and personal discussion, is constructed.
Its static geometrical properties, including degree distribution,
weight distribution, weight per degree, and betweenness
centrality, give a nice overall description of the research works.
The way we introduced here to measure the weight of connections
can be used as a general one to construct weighted network.
\end{abstract}


\section{Introduction}\label{intro}
Econophysics is a new area developed recently by the cooperation
between economists, mathematicians and physicists (See Refs.
\cite{stanley,EconRev1,EconRev2} as reviews). More and more
researchers in finance take up statistical physics to explore the
dynamical and statistical properties of financial data, including
time series of stock prices and exchange rate and size of
organizations. And also more and more physicists from statistical
physics and complexity turn to working in finance, as an important
research subject. It is interesting to investigate the
corresponding research development in this new born scientific
research area, to know the work status and to understand the idea
transportation among scientists.

Actually scientific collaboration has already become an
interesting subject for network research\cite{newman1,sci2}. So
after collecting the data and giving some basic statistical
results, we have constructed a network to study the idea
communication in this area. Here we think all collaboration,
citation, and personal discussion are the ways for idea
transportation with different contributions. If we want to analyze
this transportation as a whole we must use different weight to
measure these different levels. So the network we constructed here
is a weighted and directed one.

The database and the simple statistical results are given in the
next section. In Section \ref{network} we constructed the network
and presented some static geometrical properties. Some concluding
remarks are given in Section \ref{remark}.

\section{Database and Simple Statistical Results}
Although there is still no clear definition of Econophysics, we
just collected related works on the following three main topics
among most researchers. The first one is fluctuation of stock
prices, exchange rates, options and goods. The second is about
firm sizes, distribution of personal wealth and income, and GDP.
The last one is network analysis of economy, such as trade web,
the import/outport relationship between countries and cities.

Econophysicists have found styled facts such as volatility
clustering, nontrivial autocorrelation characters and the
universal power law distribution in the time series of stock
prices and exchange rates. Although disagreement on the styled
facts of the firm size is still exist, some universal properties
as the Laplace distribution of the return of size fluctuation have
been verified by independent works. As for the money dynamics,
both the empirical researches and the theoretical models are based
on the traditional models in physics such as ideal gas. Therefore,
although the new discipline has not been defined and accepted
widely, from practical view, it's not hard to discriminate all
these works from traditional discussion in finance and from
traditional physical models.

Concentrating on these three main topics, we started to collect
papers from the three web sites listed below. 1.)Econophysics home
page at http://www.econophysics.org. Almost complete paper list is
created and maintained there, including papers published and
preprinted. 2.)Econophysics at Physica A at
http://www.elsevier.com/locate.econophys, where all papers
published by Physica A are collected. 3.) ISI at
http://isi.knowlege.com, the well known SCI information provider.
We collected publish information of papers published from 1992 to
4/30/2003. Then we got them from the journals one by one. At last,
we get 662 papers and totaly 556 authors from more than 20
journals. Then we extracted the times of three relationship
between every two scientists from this data set to form a file of
data record as ($S_{1}, S_{2}, x, y, z$), which means author $S_1$
collaborated with author $S_{2}$ `$x$' times, cited `$y$' times of
$S_2$'s papers and thanked $S_{2}$ `$z$' times in all $S_1$'s
acknowledgements. Here we must mention that in order to keep our
data set to be closed, we only counted the cited papers that have
been collected by our data set and just selected the people in
acknowledgements which are authors in our data set.

From the above database, we can get basic statistical results
based on authors and papers.

\subsection{Statistics on authors}
\begin{figure}[th]
\begin{minipage}[t]{0.48 \textwidth}
\centerline{\psfig{file=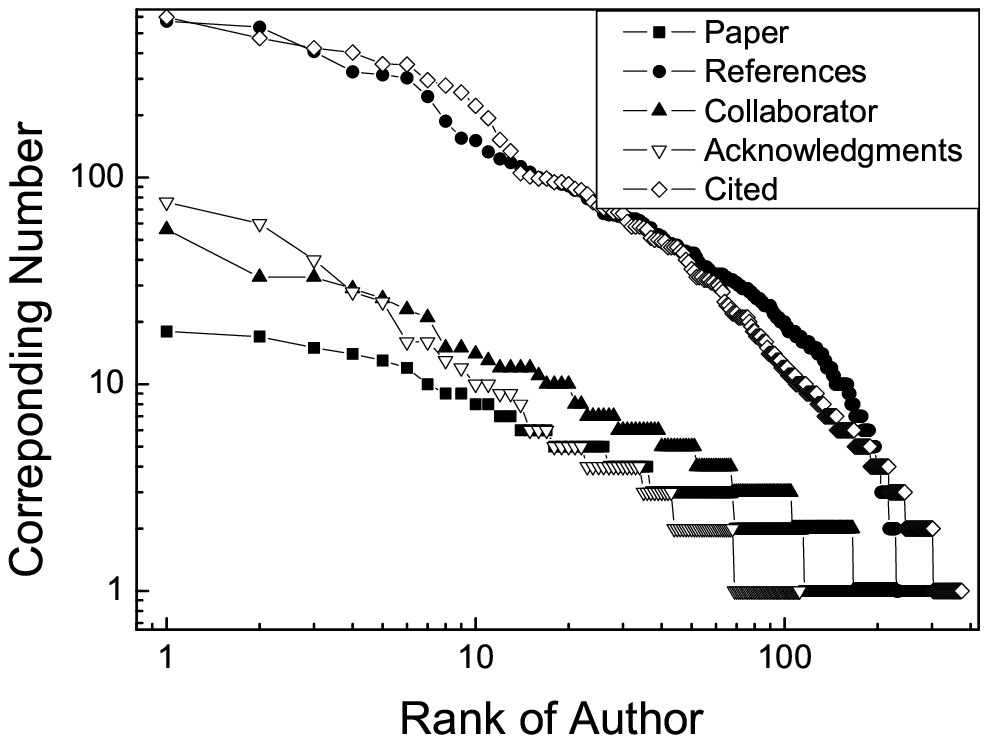, width=6cm}}
\caption{Statistics on authors} \label{author}
\end{minipage}
\begin{minipage}[t]{0.48 \textwidth}
\centerline{\psfig{file=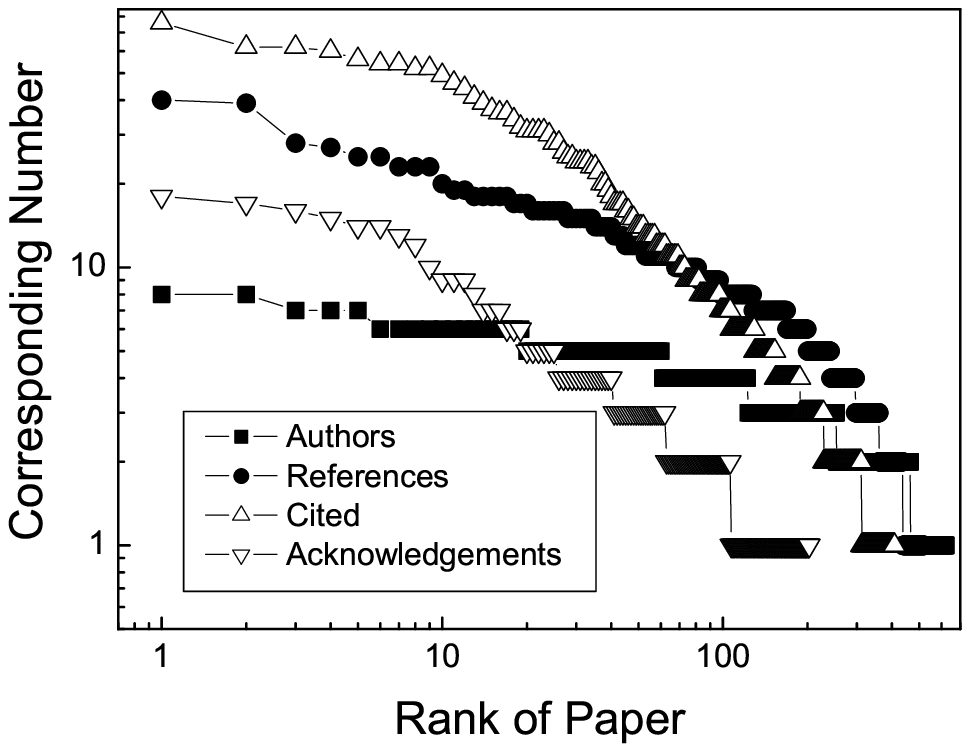, width=6cm}} \caption{Statistics
on papers} \label{paper}
\end{minipage}
\end{figure}

Every author has his/her own coauthors, papers, papers cited by
his/her papers, papers citing his/her papers, and people in all
his/her acknowledgements. All the statistical results are related
to the important properties of the communication networks. We
explored these distributions of the first author of each paper and
show them in Zipf plot\cite{zipf} as Fig. \ref{author}.
In a Zipf plot, one sorts the object number in decreasing order,
then a figure of value vs rank is plotted. This plot is equivalent
to frequency counting but is more suitable for small database and
doesn't depend on the counting interval. The coauthor number of an
author is the number of the different coauthors in all of his/her
papers. Citation times of an author is defined by the summation of
the number of references in all of his/her papers and the time
being cited is the total number of papers in which his/her paper
was cited. The acknowledgements here are thanks sending out by the
author.

\subsection{Statistics on papers}
Every paper has its own authors, references, papers citing it and
people in its acknowledgements. In this section, we also use Zipf
plot to demonstrate such distributions. Number of authors, number
of references, number of papers in which it was cited, number of
people in its acknowledgements were plotted in Fig. \ref{paper}.
All citation and acknowledgements are counted within our data set.
From Figs. \ref{author} and \ref{paper}, we can find most part is
power law distribution\cite{power}, while they are usually truncated
at high tail\cite{truncatpower}. To get universal property,
investigation on more data set is required.

\section{Network Construction and Static Geometrical Properties}\label{network}
Using the author as vertex and communication among
them as links, we can construct a scientific communication network. It is
a weighted and directed network as discussed in Section \ref{intro}.
The weight of a link is converted from the times by the formula
\begin{equation}
w_{ij}=\sum_{\mu}w^{\mu}_{ij}, \label{tanhw}
\end{equation}
in which, $\mu$ can only take value from $\{1,2,3\}$. So
$w^{\mu}_{ij}$ is one of the three relationships---coauthor,
citation or acknowledgement and is defined as
\begin{equation}
w^{\mu}_{ij}=\tanh\left(\alpha_{\mu}T^{\mu}_{ij}\right),
\label{weight}
\end{equation}
where $T^{\mu}_{ij}$ is the times of $\mu$ relationship between
$i$ and $j$.

We think the weight will reach a limitation when the times exceeds some
value. Therefore, we use the $\it tanh$ function as the weight
function. We also assume the contributions to the weight from
these three relations are different and they can be represented by
the different values of $\alpha_{\mu}$, for which $0.7,0.2,0.1$
are used for $\alpha_{1}, \alpha_{2}, \alpha_{3}$ in this paper.
The thumb principle and the effect of different parameter values
will be discussed in other paper.

At last, equation \ref{tanhw} is a weight of similarity, it is
convenient to convert it into dissimilarity weight as
\begin{equation}
\tilde{w}_{ij} = 3\times\frac{1}{w_{ij}} \hspace{0.3cm}( if
\hspace{0.15cm} w_{ij} \neq 0 ).
\end{equation}
It's timed by $3$ because the similarity weight
$w_{ij}\in\left(0,3\right]$ and timing by $3$ will normalize $\tilde{w}_{ij}\in\left[1,\infty\right)$.

\subsection{Degree, weight and weight per degree}
\begin{figure}[th]
\centerline{\psfig{file=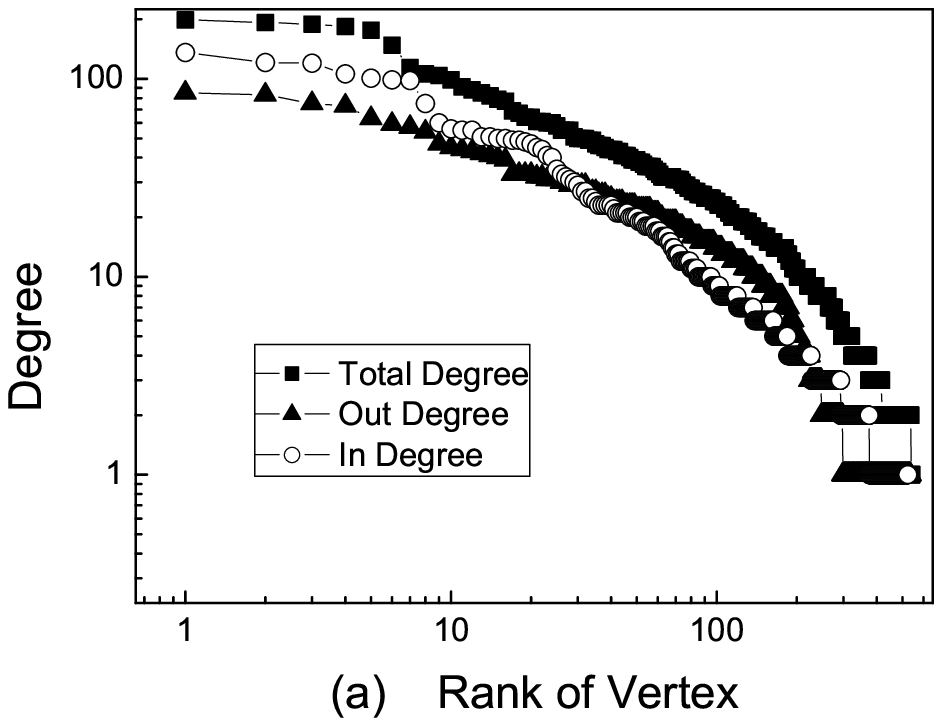,width=6cm}
\psfig{file=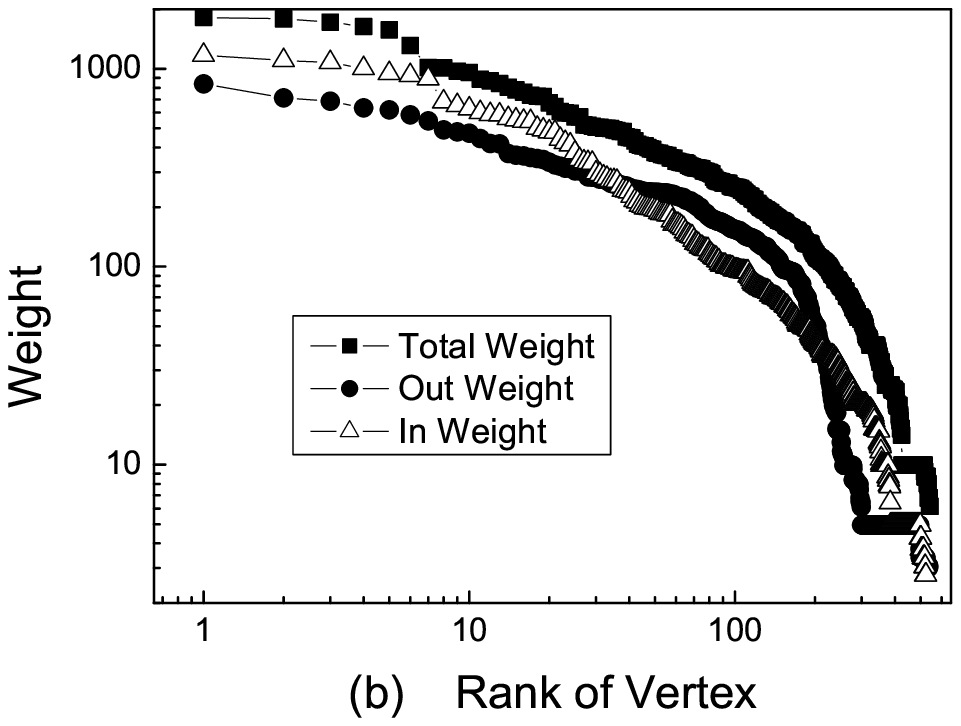,width=6cm}} \caption{Distribution of
degree(a) and weight(b)} \label{degreeweight}
\end{figure}

\begin{figure}[th]
\centerline{\psfig{file=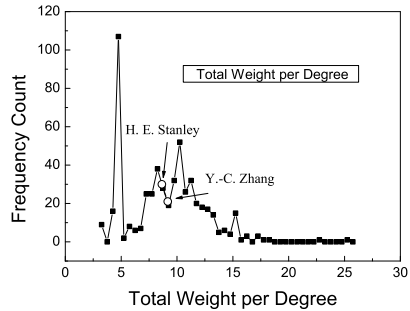,width=6cm}}
\caption{Distribution of weight per degree} \label{wpd}
\end{figure}

Since this scientists network is directed, we have three
definitions of degree: out degree, in degree and total degree. And
also three ones for weight. Fig. \ref{degreeweight} shows all these
distributions. Now we introduce another interesting
quantity -- weight per degree, which is the quotient of weight and
degree. It also has three value, in, out and total. Fig. \ref{wpd}
shows their distributions. We think weight per degree is one of
the inherent variables of vertex. It has some relations with the
working style of scientists. Preferring to cooperate widely not
deeply may lead to large weight per degree, while for other
scientists maybe wish to cooperate with few people but more deeply
and this result in small weight per degree. This quantity of a
scientist can be calculated from a collection of all his/her
papers. So it's local and easy to be calculated. Maybe this
quantity can be used as a measure of the working style of a
scientist.

\subsection{Betweenness centrality}
\begin{figure}[th]
\centerline{\psfig{file=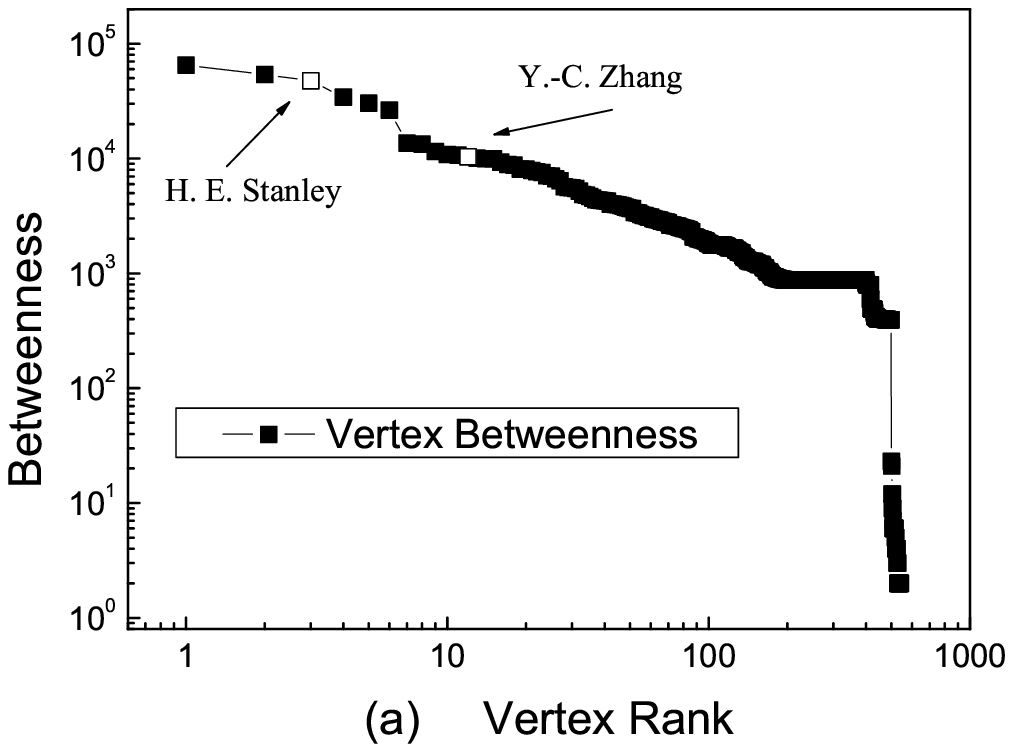,width=6.3cm}\ \   \ \
\psfig{file=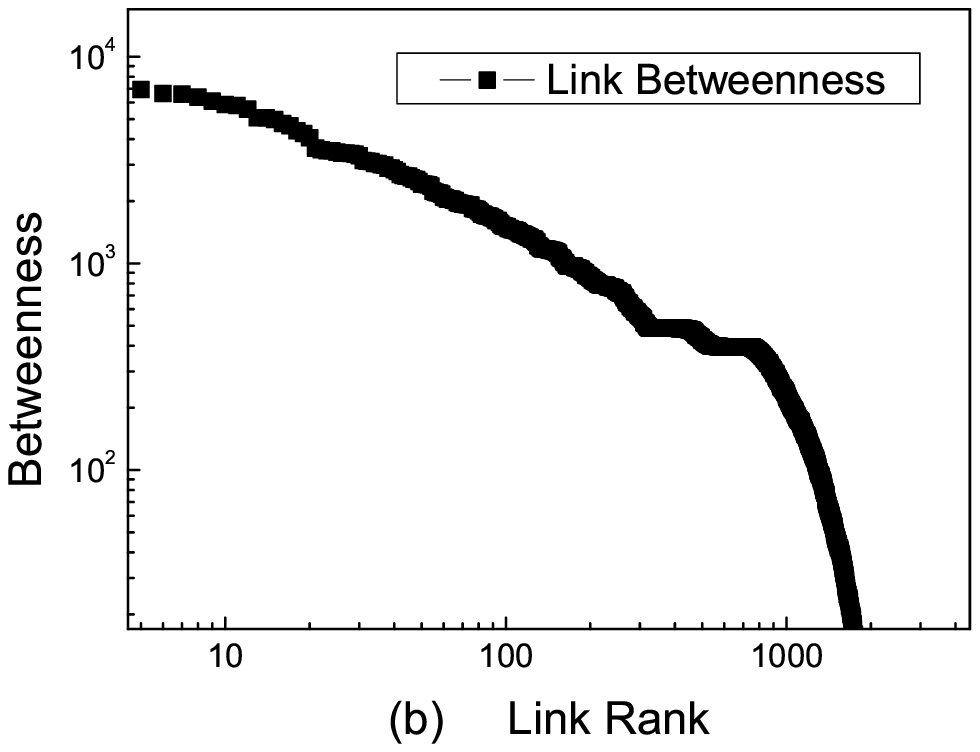,width=6cm}} \caption{Distribution of
vertex(a) and link(b) betweenness} \label{between}
\end{figure}

Other important geometrical quantities for the network are given
by average shortest distance and the vertex and link
betweenness\cite{between1,b2}. Here in Fig. \ref{between}, we give
vertex betweenness and link betweenness distributions. Betweenness
describes the relative importance of a vertex or link in the
communication in the network. A great betweenness implies it's a
key point for global communication.

\section{Concluding Remarks}\label{remark}
The development of a field is an excellent scientific object for
researchers, especially the field is a new developing topic. It's
not very hard to record all the development for such topic,
especially nowadays when we have almost complete list of all
publications. From the point view of complex networks, such
historical empirical research will definitely boost the new
exploration in mechanism of networks. Let's keep eyes on the
development of Econophysics, both on the new research works and on
the development itself.

The construction of weighted network, especially the way we
introduced to measure the weight, is also helpful for the research
on complex networks. Some further studies on this communication
network, such as effect of different measurements, clustering
coefficient and average distance for weighted network, and
transportation efficiency have been done already. Similar
discussion on a large database\cite{newman1} is in progress.
Results will be reported later.

We also wish such network analysis of development of Econophysics
can also benefit Econophysics. Some centrality analysis of vertex
reveals the relative scores of different researchers, and further
works on communication structure will show the group structure and
subject clustering information, and this maybe will imply some new
related blank topics. Such exploration will be one of our next
steps in network analysis of developing scientific fields.

\section*{Acknowledgements}
The work is supported by National Natural Science Foundation of
China under the Grant No. 70371072 and No. 70371073.

\end{document}